\documentstyle[psfig,amssymb]{mn}
 
\def\lunits{$\rm erg~s^{-1}$}

\title[Mining for normal galaxies in the 1XMM catalog]{Mining for
normal galaxies in the First XMM-Newton Serendipitous Source Catalog} 
 
\author[Georgakakis et al.] {A. E. Georgakakis$^{1,2}$\thanks{email: a.georgakakis@imperial.ac.uk},
  V. Chavushyan$^{3,4}$, M. Plionis$^{1,3}$,
  I. Georgantopoulos$^{1}$,\\ \\ 
  {\LARGE E. Koulouridis$^{1,5}$,
  I. Leonidaki$^{1,5}$, A. Mercado$^{3}$}  
  \\ \\
  $^1$Institute of Astronomy \& Astrophysics, National Observatory of
  Athens, I. Metaxa \& V. Pavlou, Athens, 15236, Greece \\ 
  $^2$Astrophysics Group, Blackett Laboratory, Imperial College, Prince
  Consort Rd , London SW7 2BZ, UK\\
  $^{3}$Instituto Nacional de Astrof\'isica \'Optica y
  Electr\'onica, AP 51 y 216, 72000, Puebla, Pue, M\'exico\\
  $^4$Instituto de Astronomia, Universidad Nacional Autonoma de
  M\'exico, A.P. 70-264, DF 04510, M\'exico\\
  $^5$Astronomical Laboratory, Department of Physics, University of
  Patras, 26500 Rio-Patras, Greece\\
}
\begin{document}
\maketitle  

\begin{abstract}
This paper uses the First {\it XMM-Newton} Serendipitous Source
Catalog compiled by the {\it XMM-Newton} Science Center to identify
low-$z$ X-ray selected normal galaxy candidates.  Our sample covers a
total area of $\approx \rm 6\, deg^2$ to the 0.5-2\,keV limit
$\rm \approx 10^{-15} \, erg \, s^{-1} \, cm^{-2}$. A total of 23 sources
are selected on the basis of low X-ray--to--optical flux ratio $\log
f_X/f_{opt}  < -2$, soft X-ray spectral properties and optical
spectra, when available, consistent with stellar than AGN
processes. This sample is combined with similarly selected systems
from the Needles in the Haystack Survey (Georgantopoulos et al. 2005)
to provide a total of 46 unique $z \la 0.2$ X-ray detected normal
galaxies, the largest low-$z$ sample yet available. This is first used  
to constrain the  normal galaxy $\log N - \log S$  at bright fluxes
($\rm 10^{-15} -10^{-13} \, erg \, s^{-1} \,  cm^{-2}$). We estimate a
slope of $-1.46\pm0.13$ for the cumulative number counts
consistent with the euclidean prediction. We further combine our
sample with 23 local ($z \la 0.2$) galaxies from the {\it Chandra}
Deep Field North and South surveys to construct  the local X-ray
luminosity function of normal galaxies. A Schechter form provides a
good fit to the data with a break at $\rm \log L_\star =
41.02_{-0.12}^{+0.14} \rm \, erg \, s^{-1}$ and a slope of
$\alpha=-1.76\pm 0.10$. Finally, for the sample of 
46 systems we explore the association between X-ray luminosity and
host galaxy properties, such as star-formation rate and stellar
mass. We find that the $L_X$ of the emission-line systems correlates
with $\rm H\alpha$ luminosity and 1.4\,GHz radio power, both providing
an estimate of the current star-formation rate. In the case of early
type galaxies with absorption line optical spectra we use the $K$-band
as proxy to stellar mass and find a correlation of the form $L_X
\propto L_K^{1.5}$. This is flatter than the $L_X - L_B$ relation for
local ellipticals. This may be due to either $L_K$ providing a better
proxy to galaxy mass or selection effects biasing our sample
against very luminous early-type galaxies, $L_X 
> 10^{42} \rm erg \, s^{-1}$.   
\end{abstract}


\section{Introduction}
The launch of the {\it XMM-Newton} and the {\it Chandra} missions has
opened new opportunities in the study of the X-ray properties of normal
galaxies. Apart from targeted observations of local systems,
surveys performed by these telescopes have allowed, for the first
time, study of normal galaxies at X-ray wavelengths outside the
nearby Universe, thus opening the way for evolutionary
investigations. Stacking analysis studies using these missions for
example, have provided  constraints on the mean X-ray properties
($L_X$, hardness ratios, $L_X/L_B$) of star-forming systems over the
redshift range $z=0.1-3$ (Hornschemeier et al. 2002; Nandra et
al. 2002; Georgakakis et al. 2003a, b; Laird et al. 2004). The
stacking analysis results from the above independent studies when
combined together are consistent with luminosity evolution of the form
$\approx (1+z)^3$ at least to $z\approx1.5$ (e.g. Georgakakis et
al. 2003a, b).    

The X-ray stacking analysis provides information on the mean
properties of systems that are too X-ray faint  to be individually
detected. A major breakthrough however, of the {\it XMM-Newton} and
the {\it Chandra} missions has been the  {\it detection} of normal
galaxies at low and moderate redshifts. Bauer et al. (2002) and
Alexander et al. (2002) studied the radio and mid-infrared properties
of X-ray sources in the  Chandra Deep Field (CDF) North and argue that
systems with X-ray emission dominated by  starburst activity are
detected in this field out to $z\approx1$. Hornschemeier et al. (2003) 
demonstrated that more quiescent normal galaxies with
X-ray--to--optical flux ratio $\log f_X / f_{opt}< -2$ and a median 
redshift $z \approx 0.3$ not only constitute a non-negligible
component of the X-ray source population in the CDF-North but  are
also likely to outnumber AGNs at fluxes below $f_X(\rm 0.5 - 2 \, keV)
\approx 10^{-17} \rm \, erg \, s^{-1} \, cm^{-2}$. Norman et
al. (2004) extended the Hornschemeier et al. (2003) study and
identified over 100 starburst/quiescent galaxy candidates to
$z\approx1$ in the combined CDF-North and South albeit with optical
spectroscopy limited to a fraction of them. These authors estimate the
X-ray luminosity function of normal galaxy candidates and find
evidence for X-ray evolution of the form $\propto (1+z)^3$ to
$z\approx1$, i.e. similar to that inferred from other wavelengths
(e.g. Hopkins 2004 and references therein).  

Parallel to the studies above focusing on distant sources ($z \ga
0.3$), there have also been efforts to compile X-ray selected normal 
galaxy samples at low redshifts $z<0.2$. This is essential to
complement the deep CDF studies and to provide a comparison sample in
the nearby Universe. Georgakakis et al. (2004) and Georgantopoulos,
Georgakakis \& Koulouridis (2005) used public {\it XMM-Newton} data
overlapping with the Sloan Digital Sky Survey (SDSS; Stoughton et
al. 2002; York et al. 2000) to identify normal galaxies at a median
redshift of about 0.07 over an $\rm 11 \, deg^2$ area. This dataset, the
Needles in the Haystack Survey, has been used to constrain the galaxy
$\log N - \log S$ at bright fluxes, estimate the X-ray luminosity
function of absorption and emission line galaxies separately and to
assess their  contribution to the diffuse X-ray background under
different evolution scenarios.  A similar study using the HRI detector
onboard {\it ROSAT} has been performed by Tajer et al. (2005) aiming
to determine the number counts of low X-ray--to--optical flux ratio
systems ($\log f_X/f_{opt} < -1$) at very bright fluxes ($\rm 10^{-14}
- 10^{-12} \, erg \, s^{-1} \, cm^{-2}$). More  recently,
Hornschemeier et al. (2005) explored the X-ray properties of
$z\approx0.1$ galaxies identified in public {\it Chandra} pointings
overlapping with the SDSS. They use this sample to investigate the
link between X-ray luminosity,  host galaxy star-formation rate (SFR)
and stellar mass. These relations are consistent with a picture where
the X-ray emission  arises in a combination of low- and high-mass
X-ray binaries as well as hot gas interstellar medium.    

Despite significance progress, the number of X-ray selected normal
galaxies at low-$z$ (e.g. $z \la 0.2$) remains small hampering
statistical studies. In this paper we further expand existing samples
using as a resource the 1st {\it XMM-Newton} Serendipitous Source
Catalogue produced by the  {\it XMM-Newton} Science Center. Our aim is
to demonstrate the power of this unique and continuously expanding
database and how it can be exploited to advance our understanding of
the properties of X-ray selected normal galaxies. Among others, such a
sample can provide a tight anchor point for comparison with deeper
surveys probing on average higher-$z$, calibrate the relation between
SFR diagnostics and X-ray luminosity and improve our
understanding of the  association between host galaxy properties
(e.g. stellar mass) and X-ray emission. Throughout this paper we adopt
$\rm H_{o} = 70 \, km \, s^{-1} \, Mpc^{-1}$, $\rm \Omega_{M} =   0.3$
and $\rm \Omega_{\Lambda} = 0.7$.         

\section{Sample selection and observations}

\subsection{The X-ray data}\label{sec_xray}

In this paper we use the First {\it XMM-Newton} Serendipitous
Source Catalogue (1\,XMM; version 1.0.1) constructed by the {\it
XMM-Newton} Survey Science Centre
(SSC\footnote{http://xmmssc-www.star.le.ac.uk/}) and released on 2003
April 7th. This is a huge database comprising source detections drawn
from 585 {\it XMM-Newton}  EPIC observations made between 2000 March 1
and 2002 May 5.  

After the initial reduction of the raw data and the filtering of the
high particle background periods the resulting event files are used to
construct images in 5 energy bands: 0.2--0.5, 0.5--2.0, 2.0--4.5,
4.5--7.5 and 7.5--12.0\,keV. The initial source detection is performed
simultaneously on the 5 energy bands above using the {\sc eboxdetect}
task of SAS. The detected sources are passed on to the SAS task {\sc
emldetect} to assess their reliability and to determine various source 
parameters by fitting the instrumental point spread function
(PSF). The source position was fit simultaneously on all input images,
whereas the source count rates are left to vary independently in each 
image. {\sc Emldetect} estimates count rates corrected for vignetting,
losses due to inter-chip gaps, bad pixels/columns, events arriving
during readout times and the extended PSF. Parallel to the source
count rate the  {\sc emldetect} task also outputs the detection
likelihood for each source providing an estimate of the probability of
spurious detection. We convert count rates to flux assuming a
power-law spectral energy distribution with index $\Gamma=1.8$ and
Galactic absorption appropriate for each individual {\it XMM-Newton}
field. The index $\Gamma=1.8$ is consistent with the median hardness
ratio, $\rm HR\approx-0.6$,  of late-type galaxies in our final
sample (see section 3). Using a Raymond-Smith model with a temperature
of 1\,keV instead of a power-law, would overestimate the
flux in the 0.5-2\,keV band by only  about 4 per cent.

In this paper we concentrate on {\it XMM-Newton} fields that have the
EPIC (European Photon Imaging Camera; Str\"uder et al. 2001; Turner et
al. 2001) cameras as the prime instrument with the PN detector
operated in full-frame mode and exposure times $>7$\,ks, to avoid
observations that are too shallow to be useful for our purposes. We
consider only sources with declination $\rm DEC(J2000)  > - 10 \,
deg$ to guarantee access from northern hemisphere telescopes and
galactic latitude $\rm |b_{II}| > 20 \, deg$ to minimise contamination from
Galactic stars. We use sources detected on the 0.5--2\,keV PN images
(i.e. we do not consider the MOS1 and 2 CCDs  here) at  off-axis angles
$<14$\,arcmin, to avoid spurious sources close to the edge of the
field of view.   

Finally, in this paper we only consider fields with right ascensions,
RA(J2000)$>4$\,hours, where our spectroscopic follow-up program of normal
galaxy candidates is complete. For systems with  $\rm RA(J2000)=0-4$\,hours
optical spectroscopy is in progress. A total of 51 {\it XMM-Newton}
fields fulfil the criteria above. 

\begin{figure}
\centerline{\psfig{figure=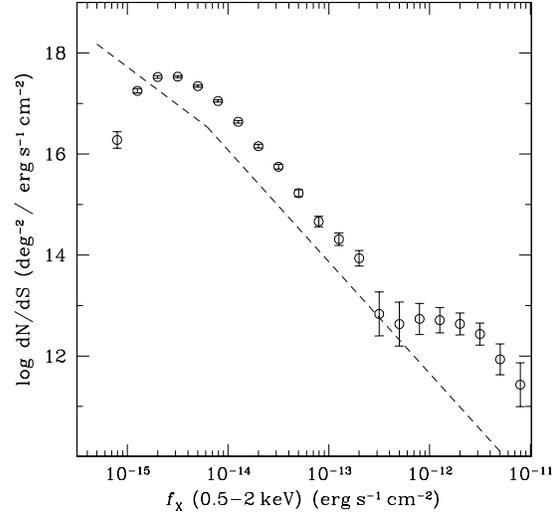,width=3in,angle=0}}
\caption
 {Differential source counts in the 0.5-2\,keV band. The dots are the
 $dN/dS$ for all the 1\,XMM catalogue  sources with likelihood
 probability $>7$ that lie on the {\it XMM-Newton} fields
 considered in this paper. The dashed line is the double power-law
 fit to the 0.5-2\,keV counts of Baldi et al. (2002). 
 }\label{fig_dnds}
\end{figure}

\begin{figure}
\centerline{\psfig{figure=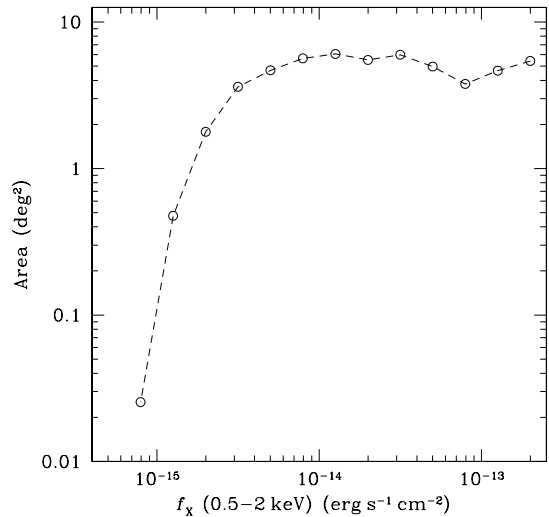,width=3in,angle=0}}
\caption
 {Solid angle as a function of limiting flux in the 0.5-2\,keV
 band.}\label{fig_area}  
\end{figure}

A drawback of the 1\,XMM catalogue is that there is no areal coverage
information providing an estimate of the surveyed area available  at
different point-source flux limits. To overpass this issue we
determine the area curve indirectly by comparing the differential
observed counts (i.e. including any incompleteness) in the 0.5-2\,keV
band with the expected $\log N - \log S$ from the literature. Here,
we adopt the double-power law parametrisation of Baldi et al. (2002)
for the 0.5-2\,keV differential source counts. For the observed counts
we use all X-ray sources in the 1\,XMM catalogue that overlap with the
{\it XMM-Newton} fields considered in this paper and 0.5-2\,keV
likelihood probability $>7$. The comparison between our uncorrected
counts and the Baldi et al. (2002) double-power law is shown in Figure
\ref{fig_dnds}. The resulting area curve is plotted in Figure
\ref{fig_area}. This is increasing to about $\rm
6\,deg^2$ at  $\approx 10^{-14} \, \rm erg \, s^{-1} \, cm^{-2}$ and
then remains almost constant.  At bright fluxes, $\ga 10^{-13} \, \rm
erg \, s^{-1}  \, cm^{-2}$, the uncorrected differential counts in
Figure \ref{fig_dnds} are affected by (i) sources associated with the
prime target of a given {\it XMM-Newton} pointing (Galactic stars or
QSOs), (ii) small number statistics and (iii) uncertainties arising
from the extrapolation of the Baldi et al. (2002) source counts to
these fluxes. As a result the area curve above $\approx 10^{-13} \, \rm
erg \, s^{-1}  \, cm^{-2}$ is unreliable. However, this does not
affect our results since all our normal galaxy candidates are fainter
than   $10^{-13} \, \rm erg \, s^{-1}  \, cm^{-2}$ (see section 3). We
also confirm that the area curve in Figure \ref{fig_area} is robust
using the Needles in the Haystack Survey (Georgantopoulos et
al. 2005). For that sample we derive the area curve using the method
outlined above and compare it with that estimated using the
sensitivity maps. We find excellent agreement between the two methods
suggesting that the area curve in Figure \ref{fig_area}  is reliable.   

\begin{figure}
\centerline{\psfig{figure=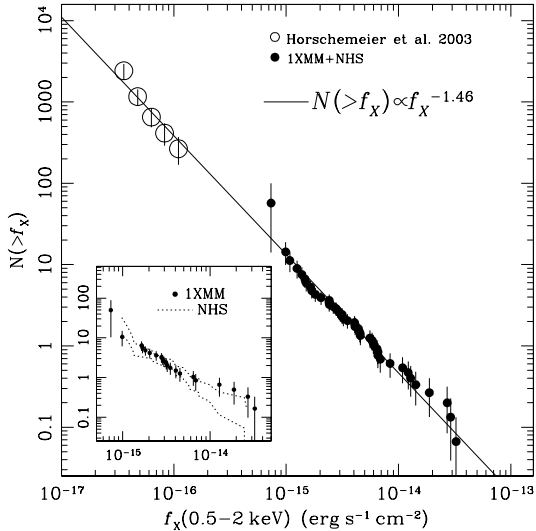,width=3in,angle=0}}
\caption
 {Cumulative normal galaxy counts in the 0.5-2\,keV spectral
 band. Filled circles are the  combined sample of normal galaxy
 candidates from the Needles in the Haystack Survey  (NHS;
 Georgantopoulos et al. 2005) and the present 
 study. The open circles are  the   source counts of Hornschemeier et
 al. (2003). The continuous line is 
 the best fit to the $\log N - \log S$ at bright fluxes from the
 combined NHS and 1XMM samples. The inlet plot shows the $\log N -
 \log S$ for the galaxies in the present study only, in comparison with
 the counts from the NHS
}\label{fig_lognlogs}    
\end{figure}

\subsection{Normal galaxy selection}
The SSC cross-correlates the 1\,XMM X-ray source positions with
various astronomical catalogues including the U.S. Naval Observatory
Catalog (USNO) version A2.0. This provides a homogeneous platform
for the optical identification of 1\,XMM sources. The USNO-A2.0
catalogue is based on photographic material and provides accurate
positions (typical error of $\approx 0.25$\,arcsec) and reasonable
$B$-band photometry with typical photometric errors of about 0.25\,mag
(Monet et al. 2003). This photometric uncertainty is adequate for this
study that identifies `normal' galaxy candidates on the basis of the
low X-ray--to--optical flux ratio $\log f_X / f_{opt} < -2$.  This
is estimated from the 0.5-2\,keV flux $f_X ( \rm 0.5 - 2 \,
keV)$ and the  USNO-A2.0 $B$-band magnitude according to the relation   

\begin{equation}\label{eq1}
\log\frac{f_X}{f_{opt}} = \log f(0.5-2\,{\rm keV}) +
0.4\,B + 5.29.
\end{equation}

\noindent
The equation above is derived from the X-ray--to--optical flux
ratio definition of Stocke et al. (1991) that involved 0.3-3.5\,keV
flux and $V$-band magnitude. These quantities are converted to
0.5-2\,keV flux and $B$-band magnitude assuming a mean colour
$B-V=0.8$ and a power-law X-ray spectral energy distribution with
index  $\Gamma=1.8$. The $\log f_X / f_{opt} = -2$ cutoff although
minimises the AGN contamination may exclude from the sample X-ray
ultra-luminous normal galaxies such as powerful  starbursts
(e.g. NGC\,3256;  $\approx 2\times10^{42}$\lunits; Moran, Lehnert \&
Helfand 1999) or very massive ellipticals ($L_X>10^{42}$\lunits;
e.g. O'Sullivan, Forbes \& Ponman 2001). Georgantopoulos et al. (2005)
argue that in the case of  star-forming galaxies the $\log f_X /
f_{opt} = -2$ limit is not likely to be a major source of
incompleteness, at least for the low-$z$ Universe. For early type
galaxies we use the local E/S0 sample of Fabbiano et al. (1992) to
estimate incompleteness because of the $\log f_X / f_{opt} = -2$
cutoff.  We find that about 20 per cent of the galaxies in that sample
have $\log L_X /L_B  >-2$, most of which are also luminous with  $L_X  
\ga 10^{42} \rm \, erg \, s^{-1}$. We note however, that some AGN
contamination is expected within this X-ray bright subsample.

The $\log f_X / f_{opt} < -2$ regime is also populated with  Galactic 
stars. The USNO A2.0 catalogue does not discriminate between  
optically extended and point-like sources. For this purpose we use
the APM scans of the UKST
plates\footnote{http://www.ast.cam.ac.uk/$\sim$apmcat} that provide 
star/galaxy separation. We further quantify the reliability
of the APM star/galaxy classification below, using follow-up
spectroscopic observations or CCD quality data from the 3rd release of
the Sloan Digital Sky Survey (SDSS DR3), where available.  

A total of 102 sources in our 1\,XMM sample have $\log f_X / f_{opt} <
-2$. APM classifies 69 of them point-like and 30 extended
(e.g. galaxies) with the remaining 3 having ambiguous classification. 
The latter class is for objects that are assigned different types on
the red and  blue UKST survey plates.   

\subsection{Optical spectroscopic observations}
 
A total of 21 out of 102 sources in our 1\,XMM sample with $\log f_X /
f_{opt} < -2$ have been observed, 5 of them are unresolved in APM with
the remaining 16 classified galaxy-like. The 5 point-like sources are
included in the target list for follow-up spectroscopic observations
to explore the reliability of the APM star/galaxy separation.  

Low resolution optical spectroscopy was carried out with the Mexican
2-m class telescopes of the Observatorio Astrofisico Guillermo Haro
(OAGH) in Cananea and Observatorio Astronomico Nacional de San Pedro
Martir (OAN SPM). Observations with the 2.1-m OAGH telescope were
carried out  with the Boller \& Chivens (B\&Ch) spectrograph and
Landessternwarte Faint Object Spectrograph and Camera (LFOSC)
Zickgraf et al. (1997).

The B\&Ch spectrograph  uses a Tektronix TK1024AB CCD mounted at the
Cassegrain focus giving a pixel scale of approximately
0.45\,arcsec. We use a 2.5\,arcsec  wide slit and a  grating with
150\,lines/mm providing a dispersion of  $\rm 3.5\,\AA\,pixel^{-1}$
and a wavelength resolution of $\approx \rm 15\,\AA$   ($\simeq
4$\,pixels FWHM)  over the range 3800--7100\,\AA.  The LFOSC is attached
to the Cassegrain focus and is equipped with an EEV P8603 385x578 CCD
giving a 10x6\,arcmin$^2$ field of view and an image scale of
1\,arcsec.  We used  a 3\,arcsec wide slit and the G3 grism giving a
wavelength resolution of  $\approx \rm 18\,\AA$ in the range
4200-9000\,\AA. The observations were carried out  during various
observing runs between December 2003 and March 2005.  

The observations with the 2.1-m OAN-SPM telescope were carried out
with the B\&Ch spectrograph equipped with SITe3 (1024x1024 pix) CCD
installed at the Cassegrain focus giving  a pixel scale of
approximately 1\,arcsec. We use a 2.5 arcsec wide slit and a grating
with 300 lines/mm, providing a dispersion of 4.5\,\AA/pix and an
effective instrumental  spectral resolution of about 10\,\AA\ ($\simeq
2$\,pixels FWHM) in the wavelength  range 3800--8000\,\AA. The 
observations were carried out during  two observing runs in November
2003 and February 2004. 

Typical exposure times for both the B\&Ch and the LFOSC were 1\,h per target. 
The observations were reduced following standard  procedures using IRAF
tasks resulting in wavelength and flux calibrated spectra. Redshifts
were determined  by visual inspection. Optical spectra of the galaxies
from these observations are presented in Appendix 1. Our own follow-up
spectroscopy is complemented with publicly available data from either
the SDSS or the literature.    

Out of the 30 sources classified extended by APM one is the prime
target of the {\it XMM-Newton} pointing (NGC\, 3184) and one has
unresolved optical light profile in the SDSS CCD quality data and
therefore is most likely associated with a Galactic star. Of the
remaining 28 sources a total of 26 are assigned spectroscopic
redshifts from either our own campaign (17) or the literature
(9). Two sources in the sample have no redshifts.

From the sources classified optically unresolved by the APM a total of
18 have either spectroscopic data from our own campaign or CCD quality
star/galaxy separation from the SDSS. All these sources are indeed
confirmed to be Galactic stars suggesting that the APM classification
is reliable. Finally, all 3 ambiguous sources are associated with
Galactic stars on the basis of the optical spectra or SDSS photometric
data.

\section{The sample}
In the analysis that follows we exclude all Galactic stars on the
basis of either optical spectroscopy or the APM/SDSS star/galaxy
separation. Our final normal galaxy sample comprises a total of 28
sources, 26 spectroscopically identified galaxies and 2 sources
without spectra classified extended by APM. This is presented  in
Table 1. For completeness we also show the 4 sources with extended (1)
or ambiguous (3) APM classification that turned out to be Galactic
stars.   

A number of sources in our sample are classified AGNs on the basis of
their optical spectroscopic properties. These include  source \#A063
which is a BAL-QSO at $z=0.149$  (Gallagher et al. 1999) and source \#A140
(NGC\,4156) suggested to harbor AGN activity by Elvis et
al. (1981). Two of the systems for which we obtained optical 
spectroscopic observations, \#A001 and A035 (see Fig. A1),  show
evidence for broad $\rm H\alpha$ emission-line with FWHM $\delta
v\approx1600$ and $2000 \rm \, km \, s^{-1}$ respectively. In both
cases the measured broad width may be partly due to the low resolution
spectrum that does not allow separation of the   $\rm H\alpha$ from
the [N\,II]\,6583\,\AA. Nevertheless, we conservatively classify these
sources AGN.  Source \#A019 has narrow optical emission lines but the
line ratios [$\rm \log ([S\,II]\,6716+31/H\alpha) \approx -0.4$, $\rm
\log ([O\,III]\,5007/H\beta) \approx +0.5$; see Fig. A1] place it in
the AGN region of the diagnostic diagrams of Kewley et al. (2001)
first introduced by  Baldwin, Phillips \& Terlevich (1981) and
Veilleux \& Osterbrock (1987).  Here we adopt the theoretical lower
bound for starburst  galaxy emission-line ratios from  Kewley et
al. (2001; see their Figure 16) to discriminate  between H\,II and AGN
dominated systems. All these sources are excluded from the analysis. 
Additionally two of our normal galaxies (sources \#A106, A149; see
Fig. A1) are Coma cluster members identified in fields targeting this
cluster and are excluded from statistical studies (e.g. $\log N - \log
S$, luminosity function).

Table 1 presents our sample including AGNs. We list:  

1. Identification number and the NED name of that source if available.

2, 3. Right ascension and declination of the optical source in J2000.

4. $B$-band magnitude from the USNO-2  version A2.0.

5. $K$-band magnitude from the 2MASS All Sky Data Release (Skrutskie
   et al. 1997). Either the 
   extended or the point source catalogue was used.

6. Offset in arcseconds between the X-ray and optical source position.

7. 0.5-2\,keV X-ray flux corrected for Galactic absorption in units of
     $10^{-14}\, \rm erg \, s^{-1} \, cm^{-2}$ .

8. Hardness ratio using the 0.5-2\,keV and the 2-4.5\,keV spectral
   bands corrected for vignetting. 

9. X-ray--to--optical flux ratio defined in equation \ref{eq1}.

10. 1.4\,GHz radio flux density in mJy from either the FIRST (Becker et
   al. 1995; White et al. 1997) or
   the NVSS (Condon et al. 1998) surveys. 

11. Redshift of the source. In the appendix we present the optical
    spectra of the normal galaxy candidates in the 1XMM sample obtained at
    the OAGH and OAN-SPM Mexican telescopes as part of this project. 

12. 0.5-2\,keV X-ray luminosity in units of $\rm erg \, s^{-1}$. For
   the k-correction a power-law spectral energy  distribution was
   adopted with photon index $\Gamma=1.8$.  

13. $\rm H\alpha$ luminosity in units of $\rm erg \, s^{-1}$. This is
measured from the optical spectra after correcting for intrinsic dust
obscuration using the Balmer decrement $\rm H\alpha / H\beta $. For
more details see section \ref{sec_correlation}. 

14. 1.4\,GHz radio luminosity.   For
   the k-correction a power-law spectral energy  distribution was
   adopted with spectral index $\alpha=0.8$.  

15. Classification on the basis of the optical spectroscopic
    observations: H\,II for starforming galaxies, ABS for absorption
    line spectra and AGN for systems showing evidence for central
    black hole accretion. There is no spectrum available for source
    \#142 with the redshift estimate, showing both emission and
    absorption line features, coming from Arp (1977). In the analysis
    that follows this is assumed to be H\,II type system.

16. APM star/galaxy separation.

\begin{figure*}
\centering
\psfig{figure=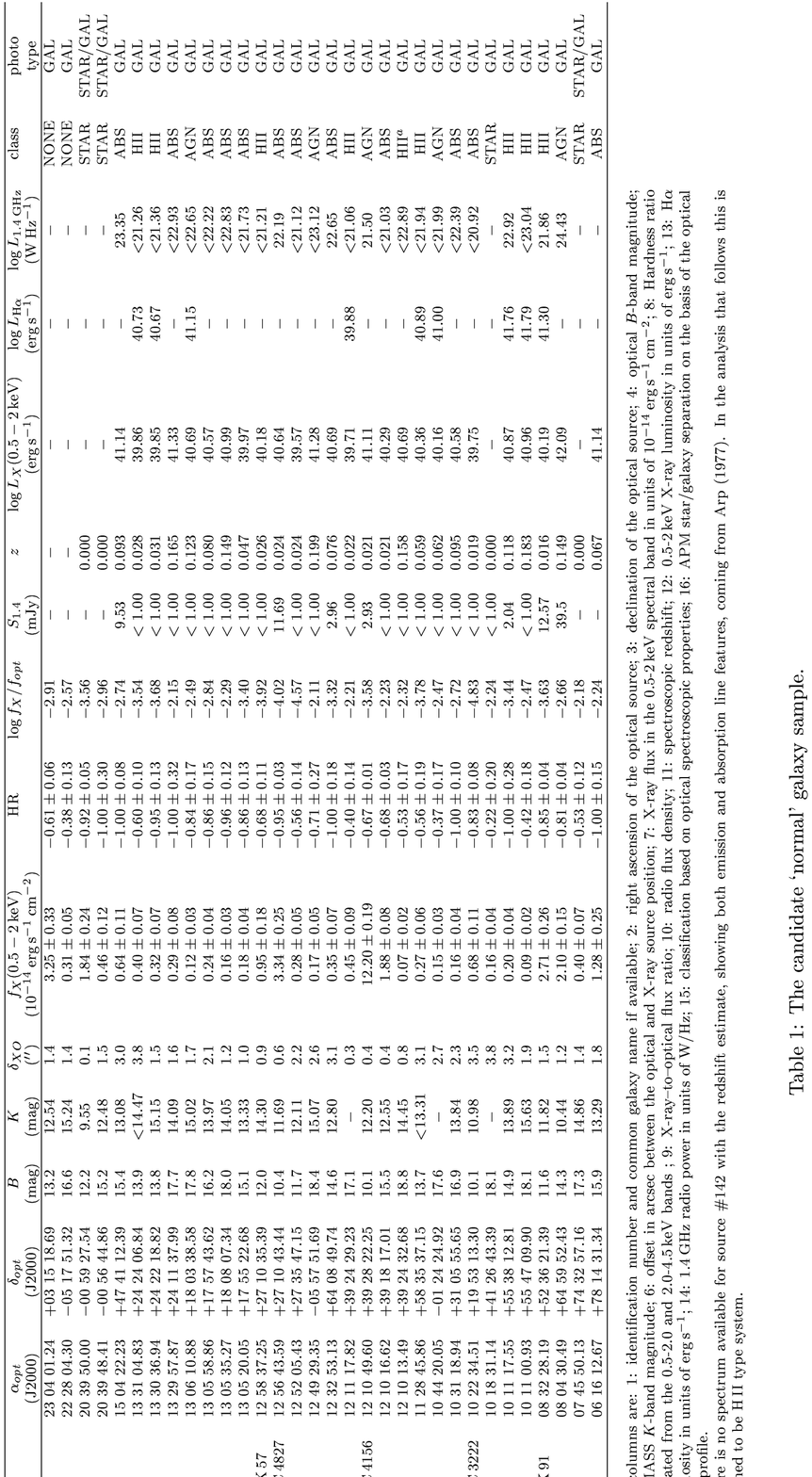,width=0.9\textwidth,angle=-180}
\end{figure*}

\section{The $\log N - \log S$} 
The inlet plot in Figure \ref{fig_lognlogs} presents the $\log N -\log
S$ in the 0.5-2\,keV band for our normal galaxy candidates derived
using the area curve from section \ref{sec_xray}. The 5 AGNs and the two
Coma cluster members in Table 1 are not used in this plot. At 
faint fluxes ($\la 10^{-14}\, \rm erg \, s^{-1} \, cm^{-2}$) there is
fair agreement with the results from the Needles in the Haystack
Survey   (NHS; Georgakakis et al. 2004; Georgantopoulos et al. 2005)
that used normal  galaxies selected in the 0.5-8\,keV band. The NHS
cumulative counts are shifted to the 0.5-2\,keV energy range assuming
$\Gamma=1.8$. At the  bright end our counts are elevated compared to
this survey, although  still  consistent within   the $1\sigma$
uncertainties. This may be because of field-to-field variations
affecting primarily  the bright flux counts derived from 
our relatively small area survey ($\rm \approx 6 \, deg^{2}$).
Despite these uncertainties at the bright end, the agreement of
our $\log N -\log S$ with previous studies suggests that the area
curve derived in section \ref{sec_xray} is robust.  

We improve the statistical reliability of our results by merging our
normal galaxy sample with that from the NHS presented by
Georgantopoulos et al. (2005). As already mentioned this sample is 
compiled in the 0.5-8\,keV spectral band using public {\it XMM-Newton}
fields overlapping with the SDSS DR-2. A total of  28 systems over
$\approx 11 \, \rm deg^2$  are identified. We shift the NHS 0.5-8\,keV
fluxes and area curve into the 0.5-2\,keV band adopting
$\Gamma=1.8$. A total of 13 overlapping fields in the two surveys 
are taken into account only once in the area curve calculation. The
combined sample covers about $\approx 15 \, \rm deg^2$ and comprises a 
total of 46 unique entries. The $\log N - \log S$ is presented in
Figure  \ref{fig_lognlogs}. The differential `normal' galaxy counts in
the 0.5-2\,keV of the two combined surveys are fit by a power law  
yielding a slope of $-2.46\pm0.13$. This correspond to a slope of
$-1.46\pm0.13$ for the cumulative number counts, in agreement with 
the results of Tajer et al. (2005) at bright fluxes and Hornschemeier
et  al. (2003;  $-1.46^{+0.28}_{-0.30}$) at the faint end, the latter
sample based on the same selection criteria used here ($\log f_X /
f_{opt}<-2$).

\section{The luminosity function} 
We further explore the statistical properties of the normal galaxy
sample by deriving the X-ray luminosity function (LF) using methods that 
are fully described by Georgantopoulos et al. (2005). We estimate 
both the binned X-ray LF using the technique of Page \& Carrera (2000)
and the parametric Maximum Likelihood fit (Tammann, Sandage \& Yahil
1979) adopting a Schechter (1976) form for the luminosity function. 

\begin{figure}
\centerline{\psfig{figure=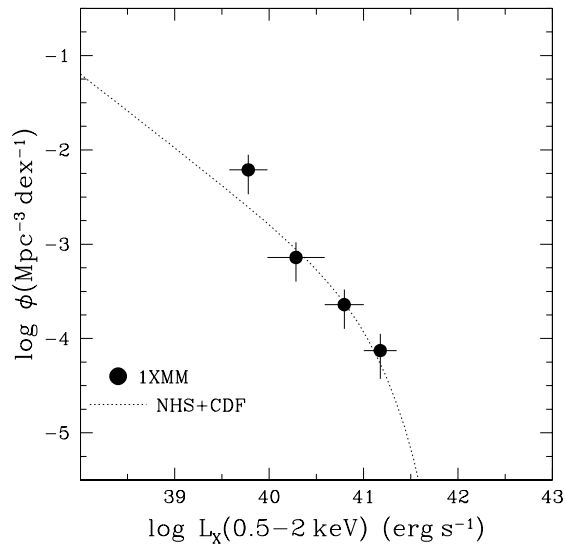,width=3in,angle=0}}
\caption
{
 The local 0.5-2\,keV luminosity function for the present sample of
 1XMM galaxies. Also shown is the maximum likelihood fit method for
 $z<0.2$ normal galaxies from the combined NHS  and CDF samples
 (see Georgantopoulos et al. 2005) shifted from the 0.5-8 to the
 0.5-2\,keV band assuming $\Gamma=1.8$.  
 }\label{fig_lf_comp}
\end{figure}

We first estimate the 0.5-2\,keV LF for the 1XMM data alone. Figure
\ref{fig_lf_comp} plots our results in comparison with those from the
combined  NHS and CDF data from Georgantopoulos et al. (2005) shifted
to the 0.5-2\,keV band assuming $\Gamma=1.8$. There is good agreement
within the $1\sigma$ errors further suggesting that the area curve
derived in section \ref{sec_xray} is adequate for statistical
studies. We note that the two sources without spectroscopic redshift
estimate are not used in this calculation. 

\begin{figure}
\centerline{\psfig{figure=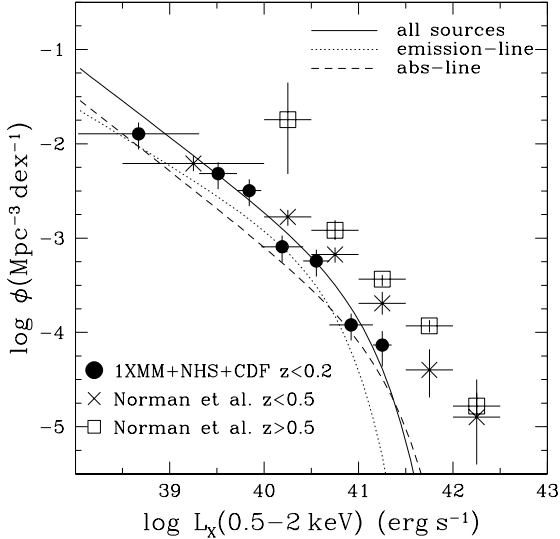,width=3in,angle=0}}
\caption
{
 The 0.5-2\,keV luminosity function for the combined sample of 1XMM,
 NHS and CDF $z<0.2$ normal galaxies. The derived maximum-likelihood
 fits to the observations are shown with the continuous (full
 sample), dotted (emission-line objects) and dashed (absorption-line
 systems) lines. For clarity the results from the non-parametric
 method are shown with filled circles only for the full sample. The
 luminosity function derived by Norman et al. (2004)  
 in two redshift bins is plotted for comparison. 
 }\label{fig_lf_all}
\end{figure}

We improve the statistical reliability of our LF estimates by
combining our sample with both the NHS  and 23 $z \la 0.2$ galaxies
selected in the 0.5-2\,keV spectral band  from the CDF-N and
CDF-S. The CDF-N data are obtained from Hornschemeier et al. (2003) by
selecting a total of 15 sources with 0.5-2\,keV band detection. All of
these systems have spectroscopic redshifts available. In the case of
the CDF-S we select a total of 8 sources detected in the 0.5-2\,keV  
spectral band with $\log f_X /f_{opt}<-2$ from the catalogue presented
by Rosati et al. (2001). Spectroscopic (total of 5) or photometric
(total of 3) redshifts are available from Szokoly et al. (2004) and
Zheng et al. (2004) respectively. This combined sample is also split
into absorption and emission line systems to explore the LF for
different galaxy types. Our results are plotted in Figure
\ref{fig_lf_all} with the maximum likelihood best-fit parameters
presented in Table  \ref{tab_xlf}, where $L_\star$ is the break of
Schechter function, $\alpha$ is the faint-end slope and $\phi_\star$
the normalisation.  In the same table we give the X-ray
emissivity (luminosity per $\rm Mpc^3$) estimated by the relation
$j_x=\int \Phi(L)~L~dL$, where $\Phi(L)$ is the luminosity function.  
The $1\sigma$ errors on $L_\star$ and
$\alpha$ are estimated from the regions around the best fit where the
likelihood function changes by $\delta L=0.5$ (e.g. Press et
al. 1992). The uncertainties in $\phi_\star$ and $j_X$ are
approximated by performing 200 bootstrap resamples of the data and
then estimating the 68th percentiles around the median. For a 
Gaussian distribution these correspond to the 68 per cent confidence
level.  

Figure \ref{fig_lf_all} also compares the X-ray luminosity function of
the combined 1XMM+NHS+CDF sample at a mean redshift $z=0.087$ with the 
higher-$z$ LF estimates of Norman et al. (2004). These authors used
Bayesian statistical analysis to select normal galaxy candidates in
the CDF-North and South and derived the first ever X-ray luminosity
function for these systems in two redshift bins with medians  $z=0.26$
and $z=0.66$  respectively.  Their $z<0.5$ LF in Figure
\ref{fig_lf_all} is in fair agreement with ours at the faint end but
diverges at brighter luminosities. This may suggest (i) contamination
of the Norman et al. (2004) sample  by AGNs at bright luminosities,
(ii) bias in our sample against X-ray ultra-luminous systems
(e.g. $L_X \ga 10^{42} \rm \, erg  \, s^{-1}$) because
of the $\log (f_X /f_{opt})<-2$ selection  or (iii) evolution of the
normal galaxy  luminosity function. In the latter case, the median
redshift of the $z<0.5$ subsample of Norman et al. is
$z_{median}=0.26$ higher than our mean of $z=0.087$. For luminosity
evolution of the form $(1+z)^{2.7}$ derived by Norman et al. (2004), a
source at $z=0.26$ is expected to become about 1.5 times more luminous
relative to $z=0.087$. Such a brightening can indeed, partially account
for the observed differences.

In Table \ref{tab_xlf} we also estimate a 0.5-2\,keV X-ray
emissivity of $(0.24\pm0.02)\times10^{38}$ and $(0.22\pm0.03) \times
10^{38} \rm \, erg \, s^{-1} \, Mpc^{-3}$ for emission and absorption line
systems respectively. These values are lower than previous
estimates. For example Georgakakis et al. (2003b) used stacking
analysis to estimate the mean X-ray properties of 2dF galaxies at
$z\approx0.1$. These authors estimate $j_X = (0.4 \pm
0.3)\times10^{38}$ and $(1.2 \pm 0.4) \times 10^{38} \rm \, erg \,
s^{-1} \, Mpc^{-3}$ for late and early type galaxies
respectively. Georgantopoulos, Basilakos \& Plionis (1999) convolved
the local optical luminosity function of the 
Ho, Filippenko  \& Sargent (1997) sample with the corresponding
$L_X-L_B$ relation  based on {\it Einstein} data. Their $j_X$
estimates scaled to the 0.5-2\,keV band are $(0.50\pm0.06)\times
10^{38} {\rm  erg\,sec^{-1}\,Mpc^{-3}}$ for H\,II galaxies and
$(0.53\pm0.06)\times 10^{38} {\rm erg\,sec^{-1}\,Mpc^{-3}}$ for
passive galaxies. The origin of the discrepancy between the emissivity
estimates presented here and those found in the studies above is not
clear. In the case of stacking analysis for example, luminous systems
may bias the mean signal toward high values, resulting in an
overestimation of the $j_X$. Although the method presented here for
estimating $j_X$ is direct and robust, we  note that a possible bias
in our sample against high $L_X$ systems may result in a systematic  
reduction of the $j_X$.

\setcounter{table}{1}

\begin{table*}
\footnotesize 
\begin{center} 
\begin{tabular}{l cc cc cc}
Sample    & Number of & mean  & $\log L_\star$      & $\alpha$ & $\phi_\star$ & $j_x$ \\  
          &   sources &  redshift        & ($\rm erg \, s^{-1}$) &
          & ($\rm \times  ln(10) \times 10^{-4} \,
          Mpc^{-3} \, dex^{-1}$)   & ($\rm \times 10^{38}  erg \, s^{-1}
          \, Mpc^{-3}$)\\
\hline 


1XMM/NHS/CDF   &  $67^{\star}$ & 0.087 & $41.02_{-0.12}^{+0.14}$  &  $-1.76^{+0.10}_{-0.10}$ &  
$3.4^{+2.5}_{-1.8}$ & $0.47_{-0.04}^{+0.04}$ \\

Emission       & 33 &  0.091 & $40.65^{+0.17}_{-0.14}$   & $-1.61^{+0.20}_{-0.17}$ & 
$6.0^{+7.3}_{-3.8}$ & $0.24^{+0.02}_{-0.02}$ \\

Absorption & 34 & 0.083 & $41.25^{+0.25}_{-0.18}$  &  $-1.79^{+0.13}_{-0.14}$  
&$0.9^{+1.1}_{-0.4}$ & $0.22^{+0.03}_{-0.03}$ \\
\hline
\multicolumn{7}{l}{$^{\star}$The 2 sources in the 1XMM sample without
redshift determination have been excluded from the analysis.}\\

\end{tabular}
\end{center}
\caption{The luminosity function best-fit parameters}
\label{tab_xlf}
\end{table*}

\section{X-ray luminosity and host galaxy properties}\label{sec_correlation} 
Using the 1XMM+NHS sample we explore the correlation between
0.5-2\,keV X-ray luminosity and indicators of star-formation
activity. The $\rm H\alpha$ luminosity ($L_{\rm H\alpha}$) measured 
from the optical spectra and the 1.4\,GHz radio luminosity ($L_{1.4}$)
obtained from either the FIRST or the NVSS surveys are used as proxies
to the galaxy SFR. Only X-ray selected normal galaxy candidates with
emission line optical spectra are considered here.   

\begin{figure}
\centerline{\psfig{figure=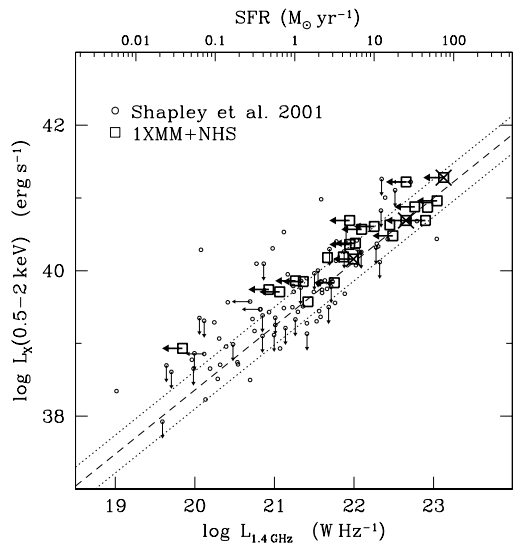,width=3in,angle=0}}
\caption
{
0.5-2\,keV X-ray luminosity against radio power. The $L_{1.4}$ is
converted into SFR on the top horizontal axis using the relation of
Bell et al. (2003). The large open squares are the emission-line
galaxies from the 1XMM+NHS sample. A cross on top of a symbol is for
AGNs. Radio upper limits are shown with an arrow pointing to the
left. The small open circles are local spiral galaxies from Shapley et 
al. (2001).  The dashed line is the best fit $L_X-L_{1.4}$ relation
of the form  $L_X \propto L_{1.4}^{+0.88}$ derived by Ranalli et
al. (2003). The dotted lines represent the 1\,sigma rms envelope
around the best fit.   
 }\label{fig_lxl14}
\end{figure}

Figure \ref{fig_lxl14} plots $L_X$ against $L_{1.4}$ for our
sources. Because of the large number of upper limits we do not attempt
to fit an $L_X - L_{1.4}$ relation to the data. Also shown in
this figure for  comparison are the spiral galaxies from  Shapley et
al. (2001). These authors compiled an optically selected sample of
local spirals with X-ray data (detections or upper limits)
from the {\it Einstein} observatory and 4.85\,GHz radio flux densities
from the literature. Powerful AGNs in this sample are not plotted in
Figure \ref{fig_lxl14}.  The {\it Einstein} X-ray fluxes in the
0.2-4\,keV band are transformed to the 0.5-2\,keV band assuming a power-law
model with a spectral index $\Gamma=1.8$. At radio  wavelengths a
power-law spectral energy distribution of the form
$f_\nu\sim\nu^{-0.8}$, appropriate for  star-forming galaxies, is
adopted to convert the 4.85\,GHz radio flux density of Shapley et
al. (2001) to 1.4\,GHz. Also shown in this plot is the $L_X - L_{1.4}$
relation  of Ranalli et al. (2003) of the form  $L_X \propto
L_{1.4}^{0.88}$. These authors  used  {\it ASCA} and {\it BeppoSAX}
data of optically selected star-forming galaxies classified  on the
basis of high quality  nuclear spectra from Ho et al. (1997). Many of
our sources are not detected at 1.4\,GHz to the 1\,mJy flux density
limit of the FIRST survey. 
The upper limits and detections, although within the scatter of the
Shapley et al. (2001) data,  appear slightly offset from the
Ranalli et al. (2003) best-fit relation, particularly for
star-formation rates $<10 \, \rm M\odot \, yr^{-1}$. This may suggest
a contribution from low-mass X-ray binaries (e.g. old stellar
population) to the observed X-ray emission of more quiescent
systems. A larger number of radio detections is required to further 
explore this issue.

\begin{figure}
\centerline{\psfig{figure=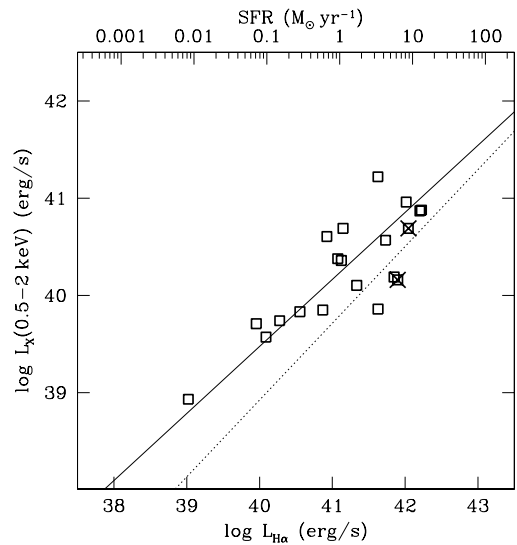,width=3in,angle=0}}
\caption
{
0.5-2\,keV X-ray luminosity against $\rm H\alpha$ luminosity corrected
for intrinsic dust obscuration for the combined 1XMM+NHS emission-line
galaxies. The $\rm H\alpha$ is converted into
SFR on the top horizontal axis using the relation of Kennicutt
(1998). A cross on top of a symbol is for AGNs. The continuous line
shows the bisector least squares fit relation of the form $L_X \propto
L_{\rm H\alpha}^{\beta}$ with $\beta=0.69\pm0.06$. The dotted line
is the $L_X - L_{\rm H\alpha}$ relation for star-forming galaxies of
Zezas (2000). Although the slopes of the two lines are similar they
differ in the normalisation. This is most likely because Zezas (2000)
used $H\alpha$  data integrated over the whole galaxy rather than slit
spectroscopy used here.  
 }\label{fig_lxlha}
\end{figure}

In Figure \ref{fig_lxlha} we plot $L_X$ against $\rm H\alpha$
luminosity for emission-line galaxies in the combined 1XMM+NHS sample
only. The latter is measured from the optical spectra after correcting
for intrinsic dust obscuration. We quantify the visual extinction,
$A_{V}$, by comparing  the H$\alpha$/H$\beta$ decrement with the
theoretical (case B recombination) value of 2.86 (Brocklehurst 1971)
and a standard reddening curve (Savage \& Mathis 1979).  The technique
is hampered by the poor S/N ratio of some of the spectra and by
stellar H$\beta$ absorption that reduces the measured flux of this
line. We account for this effect by applying a correction of
$-2$\,\AA\, to H$\beta$, similar to the mean stellar absorption in
this line determined for star-forming galaxies (Tresse et al. 1996; 
Georgakakis et al. 1999). A small number of systems in our sample have
H$\alpha$/H$\beta$ ratios below the theoretical value of 2.86 for case
B recombination, suggesting either little dust reddening, stellar
H$\beta$ absorption lower than $-2$\,\AA\, or poor signal-to-noise
ratio optical spectra. For these sources we still apply a correction
to  H$\alpha$ for dust obscuration adopting the mean optical extinction
of  $A_V=1$\,mag for  spiral galaxies determined by Kennicutt
(1992).

The best-fit relation of the form $L_X \propto L_{\rm
H\alpha}^{\beta}$ with $\beta=0.69\pm0.06$ is plotted in Figure
\ref{fig_lxlha}. This is estimated using the bisector least square
fitting described in Isobe et al. (1990). Also shown is the $L_X -
L_{\rm H\alpha}$ relation for 
star-forming galaxies determined by Zezas (2000) using a total of  43
systems with PSPC ROSAT data, classified as H\,II on the basis of high
quality nuclear optical spectra from Ho et al. (1997). The exponent of
the Zezas (2000) relation is  $\beta=0.62\pm0.11$ in good agreement
with our determination. There is however a difference in the
normalisation. This is most likely because Zezas (2000) uses $\rm 
H\alpha$ luminosities integrated over the whole galaxy whereas we are
measuring this line through a slit (or a fibre in the case of SDSS
data) with typical angular size of 2\,arcsec. This width corresponds
to a physical scale of about 2\,kpc diameter at the mean
redshift of our sample. Extrapolating fluxes measured through a slit
to integrated values over the whole galaxy is not straightforward and
depends on the size of star-forming regions, their distribution in the
galaxy, the position of the slit  and the seeing at the time of the
observations.

\begin{figure}
\centerline{\psfig{figure=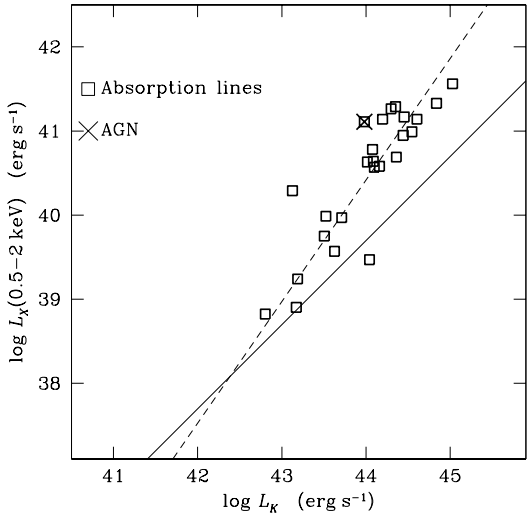,width=3in,angle=0}}
\caption
 {0.5-2\,keV X-ray luminosity against $K$-band luminosity. The large
 squares are normal galaxies in the NHS and 1XMM samples with
 absorption line optical spectra. 
 A cross on top of a symbol is for sources in our sample that show
 evidence for AGN activity.  
 The continuous line is the $L_X
 - L_K$ relation of Colbert et al. (2004) which, for a given $K$-band
 luminosity, corresponds to the mean expected contribution of the point
 source X-ray binary population to $L_X$. The dashed line is the best
 fit relation to absorption line galaxies in our combined NHS+1XMM
 sample.}\label{fig_lxlk}    
\end{figure}

We further explore the association between 0.5-2\,keV X-ray luminosity
and galaxy mass for absorption-line galaxies in the combined 1XMM+NHS
sample. The $K$-band luminosity is used as proxy to the stellar
mass of the system. Figure  \ref{fig_lxlk} plots $L_X$ against 
$L_K$. 
Also shown in this figure is the relation between the total X-ray
luminosity of point   sources, $L_{XP}$, and the host galaxy $L_K$
determined by Colbert et al. (2004) using Chandra observations of
nearby ellipticals. This line shows the mean expected contribution of
low-mass X-ray binaries to the observed emission at a given $K$-band
luminosity or galaxy mass and does not include the X-ray emitting hot
gas component.   

For absorption-line systems in our sample (excluding AGNs) we adopt a
relation of the form $L_X \propto L_K^{\gamma}$ to estimate a
bisector exponent $\gamma=1.5 \pm  0.1$. This is flatter than the
slope of $\approx1.8$ found by Fabbiano et al. (1992) for the
$L_X - L_B$ relation of local E/S0s and Hornschemeier et al. (2005)
for the $L_X$/stellar-mass correlation of absorption line galaxies in
the SDSS. This apparent discrepancy may be due to the $\log
f_X/f_{opt}<-2$ cutoff that effectively excludes from the sample 
very luminous X-ray selected normal galaxies. For example about 20 per
cent of the Fabbiano et al. (1992) sample have $\log L_X /L_B
>-2$. The vast majority of these systems are also luminous with  $L_X
\ga 10^{42} \rm \, erg \, s^{-1}$. Excluding E/S0 galaxies with $\log L_X
/L_B >-2$ from that sample we find a flatter slope of  $\approx 1.5$
for the $L_X - L_B$ relation in agreement with the value estimated
here. Alternative the flatter  $L_X - L_K$ relation may indicate that
the $K$-band provides a better proxy to galaxy mass.
For example, Shapley et al. (2001) used $H$-band near-infrared data and
find a flatter slope for the $L_X - L_H$ relation of spirals compared
to the $L_X - L_B$ relation for the same systems. Although  comparing
the X-ray properties of ellipticals and spirals is not appropriate it
suggests that the flatter slope we are estimating may be partly due to
the use of near-infrared rather than optical luminosities.  

\section{Summary and conclusions}

In this paper we demonstrate the power of the First {\it XMM-Newton}
Serendipitous Source Catalog for studies of X-ray selected normal
galaxies. Our sample is compiled from X-ray sources detected on {\it
XMM-Newton} pointings that have  (i) EPIC-PN detector as prime
instrument operated in full-frame mode, (ii) exposure time $>7$\,ks,
(iii) declinations $\rm DEC(J2000)  > - 10 \, deg$, (iv) galactic
latitude $\rm  |b_{II}|>20 \, deg$ and (v) right ascension $\rm
RA(J2000)>4$\,hours. A total of 51 fields fulfil the above criteria,
covering a total area of $\approx \rm 6\, deg^2$ to the 0.5-2\,keV
band limit  $f_X(\rm 0.5 - 2 \,  keV) \approx 10^{-15} \, erg \,
s^{-1}   \, cm^{-2}$. 

The USNO A2.0 catalogue provides a homogeneous platform that allows
identification of X-ray selected normal galaxy candidates on the basis
of the low X-ray--to--optical flux ratios of these sources, $\log f_X
/f_{opt} < -2$. Reliable star/galaxy separation is available from the
APM, allowing us to exclude Galactic stars that also have $\log f_X
/f_{opt} < -2$. Optical spectroscopy is then used to identify systems
that show evidence for AGN activity resulting in a sample of 23 normal
galaxy candidates: 9 with narrow emission lines, 12 with absorption 
lines only and 2 without optical spectroscopic information. Future  
releases of the  {\it XMM-Newton} Serendipitous Source Catalog will
provide much wider areal coverage resulting in significantly larger
low-$z$ normal galaxy samples. 

At present we increase our sample size by combining it with X-ray
selected normal galaxy candidates from the Needles in the Haystack
Survey  (Georgantopoulos et al. 2005). This provides a total of 46 $z
\la 0.2$ X-ray detected normal galaxies, the largest low-$z$ sample
yet available. Such a large number of sources provides a unique
opportunity to constrain the normal galaxy $\log N - \log S$  at
bright fluxes ($\rm 10^{-15} -10^{-13} \, erg \, s^{-1} \,   cm^{-2}$). We
estimate a slope of $-1.46\pm0.13$ consistent with the
euclidean prediction and in agreement with previous determinations
(Hornschemeier et al. 2003; Tajer et al. 2005).

Our enlarged sample is further combined with 23 local ($z \la 0.2$)
galaxies from the {\it Chandra} Deep Field North and South surveys to
construct  the local X-ray luminosity function of  normal
galaxies. This is fit with a  Schechter function with a break at $\rm  
\log L_\star = 41.02_{-0.12}^{+0.14} \rm \, erg \, s^{-1}$ and a slope
of $\alpha=-1.76\pm 0.10$. The large sample size allows us to
also estimate the luminosity function of emission and absorption line
systems separately and to provide the most accurate estimates of the
0.5-2\,keV X-ray emissivity of these systems in the nearby Universe.  
We find $j_X= (0.24 \pm 0.02) \times 10^{38} $ and $(0.22 \pm 0.03)
\times 10^{38} \rm \, erg \, s^{-1}\, Mpc^{-3}$ for emission and
absorption line systems respectively. We note that these are lower
than previous (less direct) estimates based on stacking analysis
(e.g. Georgakakis et al. 2003b) or optically selected local galaxies
(Georgantopoulos et al. 1999).     

Finally, for the combined sample of 46 systems we explore the 
association between X-ray luminosity and host galaxy properties, such
as SFR and stellar mass. We use the 1.4\,GHz radio and $\rm H\alpha$
luminosities as star-formation indicators.  We find that 
the $L_X - L_{1.4}$ relation for our sources is consistent with that
determined by Ranalli et al. (2003) for the local Universe. For the
$\log L_X - \log L_{H\alpha}$ correlation we estimate a slope of
$\approx 0.7$ in fair agreement with previous results (Zezas
2000). The above relations suggest that the X-ray luminosity in the
emission-line galaxies in our sample is directly related to the
current star-formation activity as measured by the 1.4\,GHz radio and
$\rm H\alpha$ luminosities. For the early type galaxies in the sample
showing  absorption optical lines only we use the $K$-band as proxy to 
their stellar mass and find a linear correlation between $\log L_X$
and $\log L_K$ with a slope of $\approx 1.5$. This is flatter than the 
slope of $\approx 1.8$ for the $L_X - L_B$ relation for local
ellipticals of Fabbiano et al. (1992) or the $L_X$/stellar-mass
correlation for absorption line galaxies in the SDSS determined by
Hornschemeier et al. (2005). This may be due to a possible bias in our
sample against very luminous galaxies, $L_X > 10^{42} \rm erg \,
s^{-1} \, cm^{-2}$, likely introduced by the low X-ray--to--optical
flux ratio selection. Alternatively, this may indicate that the
$K$-band provides a better representation of the galaxy mass.  

\section{Acknowledgments}
 We thank the anonymous referee for valuable comments that
 significantly improved this paper. 
 AG acknowledges funding by the European Union and the Greek
 Ministry of Development  in the framework of the programme `Promotion of
 Excellence in Technological Development and Research', project
 `X-ray Astrophysics with ESA's mission XMM'. We acknowledge the 
 use of data from the {\it XMM-Newton} Science Archive at VILSPA. 
 VC acknowledges the CONACYT research grant 39560-F.

 Funding for the creation and distribution of the SDSS Archive has
 been provided by the Alfred P. Sloan Foundation, the Participating
 Institutions, the National Aeronautics and Space Administration, the
 National Science Foundation, the U.S. Department of Energy, the
 Japanese Monbukagakusho, and the Max Planck Society. The SDSS Web
 site is http://www.sdss.org/. The SDSS is managed by the
 Astrophysical Research Consortium (ARC) for the Participating
 Institutions. The Participating Institutions are The University of
 Chicago, Fermilab, the Institute for Advanced Study, the Japan
 Participation Group, The Johns Hopkins University, Los Alamos
 National Laboratory, the Max-Planck-Institute for Astronomy (MPIA),
 the Max-Planck-Institute for Astrophysics (MPA), New Mexico State
 University, University of Pittsburgh, Princeton University, the
 United States Naval Observatory, and the University of Washington. 

\appendix
\section{}

In this appendix we present the optical spectra of the normal galaxy
candidates in the 1XMM sample obtained at the OAGH and OAN-SPM Mexican
telescopes as part of this project. 

\begin{figure*}
\centerline{\psfig{figure=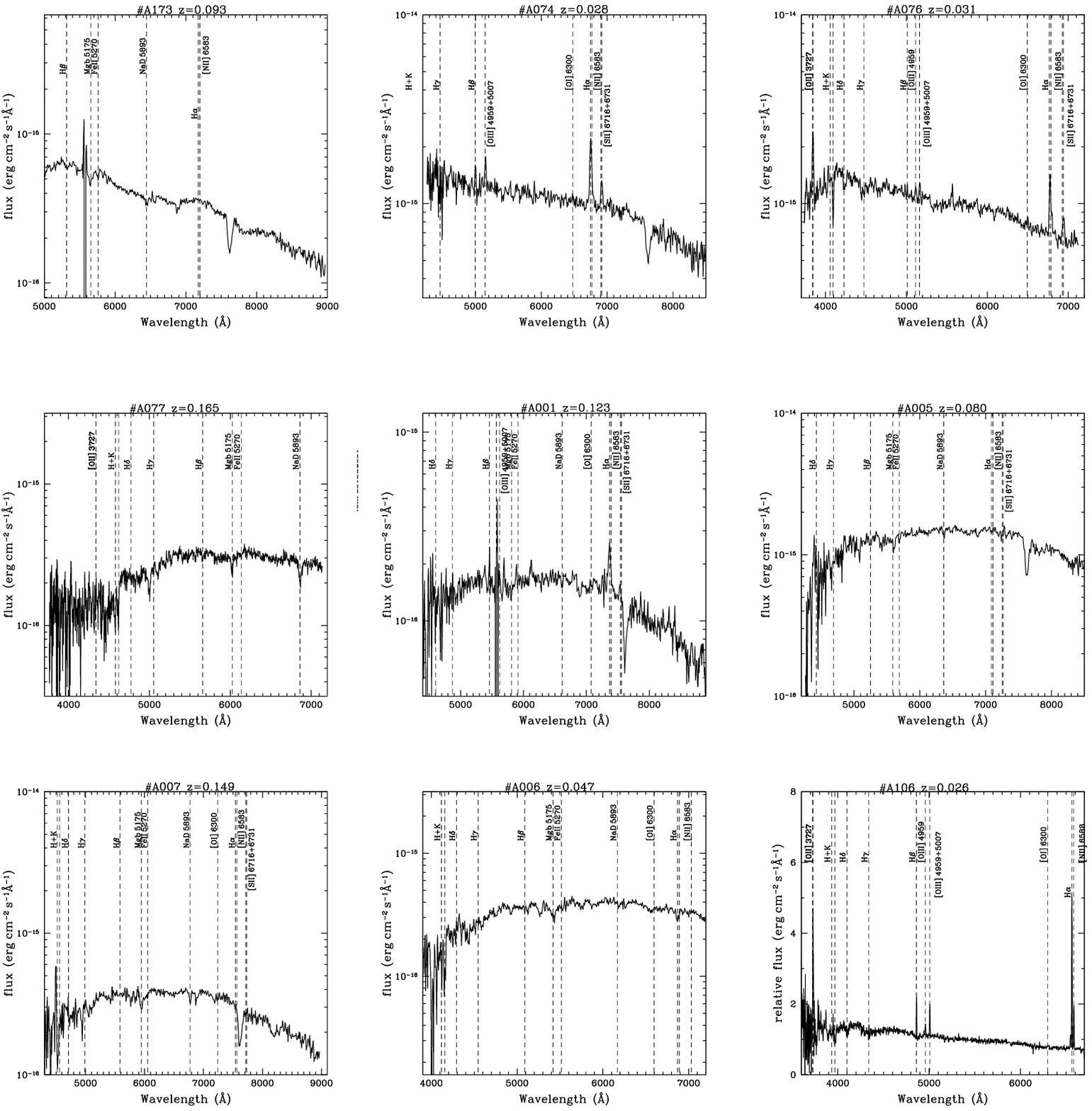,width=7in,angle=0}}
\caption
{
Optical spectra of galaxy candidates in the 1XMM sample obtained at the
OAGH and OAN-SPM Mexican telescopes. Sources A001, A035 and A019 are
classified AGN.
 }
\end{figure*}

\begin{figure*}
\contcaption{}
\centerline{\psfig{figure=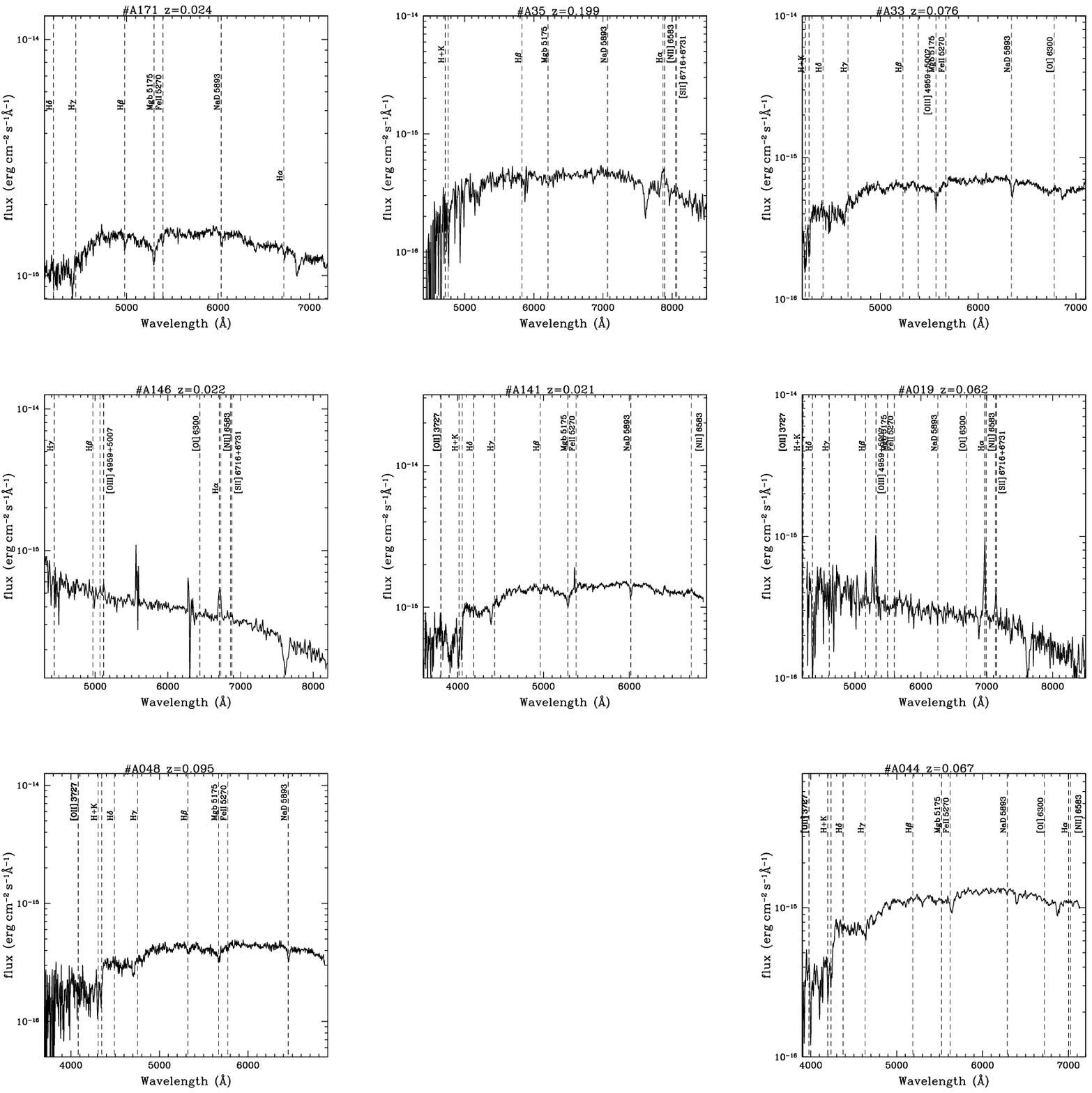,width=7in,angle=0}}
\end{figure*}

\end{document}